\documentstyle[prl,aps,graphicx,amssymb]{revtex}
\tightenlines
\makeatletter
\def\setcaption#1{\def\@captype{#1}}
\makeatother

\begin{document}


\title{ Measurement of the flux and zenith-angle distribution of upward
  through-going muons by Super-Kamiokande}

\date{\today}
\maketitle

{\center \large The Super-Kamiokande Collaboration\\}

\begin{center}
\newcounter{foots}
Y.Fukuda$^a$, T.Hayakawa$^a$, E.Ichihara$^a$, K.Inoue$^a$,
K.Ishihara$^a$, H.Ishino$^a$, Y.Itow$^a$,
T.Kajita$^a$, J.Kameda$^a$, S.Kasuga$^a$, K.Kobayashi$^a$, Y.Kobayashi$^a$, 
Y.Koshio$^a$,   
M.Miura$^a$, M.Nakahata$^a$, S.Nakayama$^a$, 
A.Okada$^a$, K.Okumura$^a$, N.Sakurai$^a$,
M.Shiozawa$^a$, Y.Suzuki$^a$, Y.Takeuchi$^a$, Y.Totsuka$^a$, S.Yamada$^a$,
%
M.Earl$^b$, A.Habig$^b$, E.Kearns$^b$, 
M.D.Messier$^b$, K.Scholberg$^b$, J.L.Stone$^b$,
L.R.Sulak$^b$, C.W.Walter$^b$, 
%
M.Goldhaber$^c$,
T.Barszczak$^d$, D.Casper$^d$, W.Gajewski$^d$,
W.R.Kropp$^d$, 
L.R.Price$^d$, F.Reines$^{d,\dagger}$, M.Smy$^d$, H.W.Sobel$^d$, 
M.R.Vagins$^d$,
%
K.S.Ganezer$^e$, W.E.Keig$^e$,
%
R.W.Ellsworth$^f$,
%
S.Tasaka$^g$,
%
\addtocounter{foots}{1}
J.W.Flanagan$^{h,\fnsymbol{foots}}$,
A.Kibayashi$^h$, J.G.Learned$^h$, S.Matsuno$^h$,
V.J.Stenger$^h$, D.Takemori$^h$,
%
T.Ishii$^i$, J.Kanzaki$^i$, T.Kobayashi$^i$, S.Mine$^i$, 
K.Nakamura$^i$, K.Nishikawa$^i$,
Y.Oyama$^i$, A.Sakai$^i$, M.Sakuda$^i$, O.Sasaki$^i$,
%
S.Echigo$^j$, M.Kohama$^j$, A.T.Suzuki$^j$,
%
T.J.Haines$^{k,d}$,
%
E.Blaufuss$^l$, B.K.Kim$^l$, R.Sanford$^l$, R.Svoboda$^l$,
%
M.L.Chen$^m$,J.A.Goodman$^m$, G.W.Sullivan$^m$,
%
%
J.Hill$^n$, C.K.Jung$^n$, K.Martens$^n$, C.Mauger$^n$, C.McGrew$^n$,
E.Sharkey$^n$, B.Viren$^n$, C.Yanagisawa$^n$,
%
W.Doki$^o$,
K.Miyano$^o$,
H.Okazawa$^o$, C.Saji$^o$, M.Takahata$^o$,
%
Y.Nagashima$^p$, M.Takita$^p$, T.Yamaguchi$^p$, M.Yoshida$^p$, 
%
S.B.Kim$^q$, 
M.Etoh$^r$, K.Fujita$^r$, A.Hasegawa$^r$, T.Hasegawa$^r$, S.Hatakeyama$^r$,
T.Iwamoto$^r$, M.Koga$^r$, T.Maruyama$^r$, H.Ogawa$^r$,
J.Shirai$^r$, A.Suzuki$^r$, F.Tsushima$^r$,
%
M.Koshiba$^s$,
%
M.Nemoto$^t$, K.Nishijima$^t$,
%
T.Futagami$^u$, Y.Hayato$^u$, 
Y.Kanaya$^u$, K.Kaneyuki$^u$, Y.Watanabe$^u$,
%
D.Kielczewska$^{v,d}$, 
%
\addtocounter{foots}{1}
\addtocounter{foots}{1} 
R.A.Doyle$^{w,\fnsymbol{foots}}$, 
\addtocounter{foots}{1}
J.S.George$^{w,\fnsymbol{foots}}$, A.L.Stachyra$^w$,
\addtocounter{foots}{1}
L.L.Wai$^{w,\fnsymbol{foots}}$, 
R.J.Wilkes$^w$, K.K.Young$^{w,\dagger}$

\footnotesize \it

$^a$Institute for Cosmic Ray Research, University of Tokyo, Tanashi,
Tokyo 188-8502, Japan\\
$^b$Department of Physics, Boston University, Boston, MA 02215, USA  \\
$^c$Physics Department, Brookhaven National Laboratory, Upton, NY 11973, USA \\
$^d$Department of Physics and Astronomy, University of California, Irvine,
Irvine, CA 92697-4575, USA \\
$^e$Department of Physics, California State University, 
Dominguez Hills, Carson, CA 90747, USA\\
$^f$Department of Physics, George Mason University, Fairfax, VA 22030, USA \\
$^g$Department of Physics, Gifu University, Gifu, Gifu 501-1193, Japan\\
$^h$Department of Physics and Astronomy, University of Hawaii, 
Honolulu, HI 96822, USA\\
$^i$Inst. of Particle and Nuclear Studies, High Energy Accelerator
Research Org. (KEK), Tsukuba, Ibaraki 305-0801, Japan \\
$^j$Department of Physics, Kobe University, Kobe, Hyogo 657-8501, Japan\\
$^k$Physics Division, P-23, Los Alamos National Laboratory, 
Los Alamos, NM 87544, USA. \\
$^l$Department of Physics and Astronomy, Louisiana State University, 
Baton Rouge, LA 70803, USA \\
$^m$Department of Physics, University of Maryland, 
College Park, MD 20742, USA \\
%
%
$^n$Department of Physics and Astronomy, State University of New York, 
Stony Brook, NY 11794-3800, USA\\
$^o$Department of Physics, Niigata University, 
Niigata, Niigata 950-2181, Japan \\
$^p$Department of Physics, Osaka University, Toyonaka, Osaka 560-0043, Japan\\
$^q$Department of Physics, Seoul National University, Seoul 151-742, Korea\\
$^r$Department of Physics, Tohoku University, Sendai, Miyagi 980-8578, Japan\\
$^s$The University of Tokyo, Tokyo 113-0033, Japan \\
$^t$Department of Physics, Tokai University, Hiratsuka, Kanagawa 259-1292, 
Japan\\
$^u$Department of Physics, Tokyo Institute for Technology, Meguro, 
Tokyo 152-8551, Japan \\
$^v$Institute of Experimental Physics, Warsaw University, 00-681 Warsaw,
Poland \\
$^w$Department of Physics, University of Washington,    
Seattle, WA 98195-1560, USA    \\
\end{center}


\begin{abstract}
  A total of 614 upward through-going muons of minimum energy 1.6~GeV
  are observed by Super-Kamiokande during 537 detector live days.  The
  measured muon flux is $(1.74\pm0.07 {\mbox{(stat.)}}\pm0.02
  {\mbox{(sys.)}})\times10^{-13} {\rm{cm^{-2}s^{-1}sr^{-1}}}$ compared
  to an expected flux of $(1.97\pm0.44(\rm{theo.}))\times
  10^{-13}{\rm{cm^{-2}s^{-1}sr^{-1}}}$.  The absolute measured flux is
  in agreement with the prediction within the errors.  However, the
  zenith angle dependence of the observed upward through-going muon flux
  does not agree with no-oscillation predictions.  The observed
  distortion in shape is consistent with
  the $\nu_{\mu}\leftrightarrow\nu_{\tau}$ oscillation hypothesis
  with $\sin^22\theta > 0.4$ and $1\times 10^{-3} <
  \Delta m^2 < 1\times 10^{-1}$ eV$^{2}$ at 90 \% confidence level.
\end{abstract}


Energetic atmospheric $\nu_{\mu}$ or $\bar{\nu}_{\mu}$ passing
through the Earth interact with the rock surrounding the
Super-Kamiokande (``Super-K'') detector and produce muons via weak
interactions.  While those neutrino-induced muons traveling downwards
are impossible to differentiate from the constant rain of cosmic ray
muons, upward-going muons are mostly $\nu_\mu$ or $\bar{\nu}_\mu$
induced, because upward-going cosmic ray muons cannot penetrate through
the whole Earth and $\nu_e$ and $\bar{\nu}_e$ induced electrons and
positrons shower and die out in the rock before reaching the detector.
Those muons energetic enough to cross the entire detector are defined as
``upward through-going muons''.  The mean energy of their parent
neutrinos is approximately 100~GeV.  Neutrinos arriving vertically
travel roughly 13,000~km, while those coming from near the horizon
originate only $\sim$500~km away.

Previously published results on atmospheric neutrinos with average
energies below $\sim$10~GeV have indicated an anomalously low
$\nu_\mu$/$\nu_e$ ratio\cite{kamsubgev,kammultigev,casper,allison} and
have also reported a strong zenith angle dependence \cite{kammultigev}.
This has been interpreted as a possible signature of neutrino
oscillations.  Recent results from this
experiment\cite{sksubgev,skmultigev} have shown strong evidence for
$\nu_\mu\leftrightarrow\nu_\tau$ oscillations\cite{skosc}.  These
results have reported on lower energy $\nu_\mu$ and $\nu_e$ neutrinos
which interacted in the water of the detector itself, hereafter referred
to as ``contained'' events.

The oscillation hypothesis has also been suggested to explain the
anomalous upward through-going muon zenith angle distributions observed
by Kamiokande~\cite{kamupthrumu} and MACRO~\cite{macro} as well as the
low absolute upward-going muon flux seen in MACRO.  However, the absolute
upward-going muon fluxes measured in Kamiokande, IMB~\cite{szendy},
and Baksan~\cite{baksan} were consistent with the no-oscillation
expectations within the large errors present in the absolute flux
predictions.

We make the first report on the measurement of upward through-going muon
flux and its zenith-angle distribution as observed by Super-K.
The experimental site is located at the Kamioka Observatory, Institute
for Cosmic Ray Research, the University of Tokyo, 1000~m underground in
the Kamioka mine, Gifu prefecture, Japan.  

The Super-K detector is a 50~kton cylindrical water Cherenkov
calorimeter.  The detector is divided by an optical barrier instrumented
with photomultiplier tubes (``PMT''s) into a cylindrical primary
detector region (the Inner Detector, or ``ID'') and a surrounding shell
of water (the Outer Detector, or ``OD'') serving as a cosmic ray veto
counter.  Details of the detector can be found in
reference~\cite{sksubgev}.

The cosmic ray muon rate at Super-K is 2.2~Hz.  The trigger efficiency
for a muon entering the detector with momentum more than 200~MeV/c is
$\sim$100\% for all zenith angles.  The nominal detector effective area
for upward through-going muons with a track length \(>\) 7m in the ID is
$\sim$1200~m$^2$.

The data used in this analysis were taken from Apr.~1996 to Jan.~1998,
corresponding to 537 days of detector livetime.  Event reconstruction is
made by means of the charge and timing information recorded by each hit
PMT.  The direction of a muon track is first reconstructed by several
automated grid search methods, which find the track by minimizing the
width of the residual distribution of the photon time-of-flight
subtracted ID PMT times.  Details of one such muon fitter are described
elsewhere~\cite{hatakeyama}.

A minimum track length cut of 7m ($\sim$1.6~GeV) was applied.  To reduce
the abundant downward-going cosmic ray muons, events satisfying
$\cos\Theta<0.1$ are selected, where $\Theta$ is the zenith angle of the
muon track, with $\cos\Theta<0$ corresponding to upward-going events.
Muons which leave both entrance and exit signal clusters in the OD are
regarded as through-going.  After a visual scan by two independent
groups (event loss probability \(<\) 0.01\%) and a final direction
hand-fit, 614 upward through-going muon events with $\cos\Theta<0$ are
observed.  Different hand fits are consistent with each other within
$1.5^\circ$.  They are shown to be unbiased in blind tests using
Monte-Carlo (MC) simulated events, with deviations between the
reconstructed track direction and the real muon direction
($\Delta\theta_{rec}$) estimated to be $1.4^\circ$.  Using this same MC,
the directional correlation between a muon and its parent neutrino is
estimated to be 4.1$^\circ$, including contributions from the muon
production angle and from multiple Coulomb scatterings in the rock.

\begin{figure}[thbp]
  \includegraphics[width=6.5in]{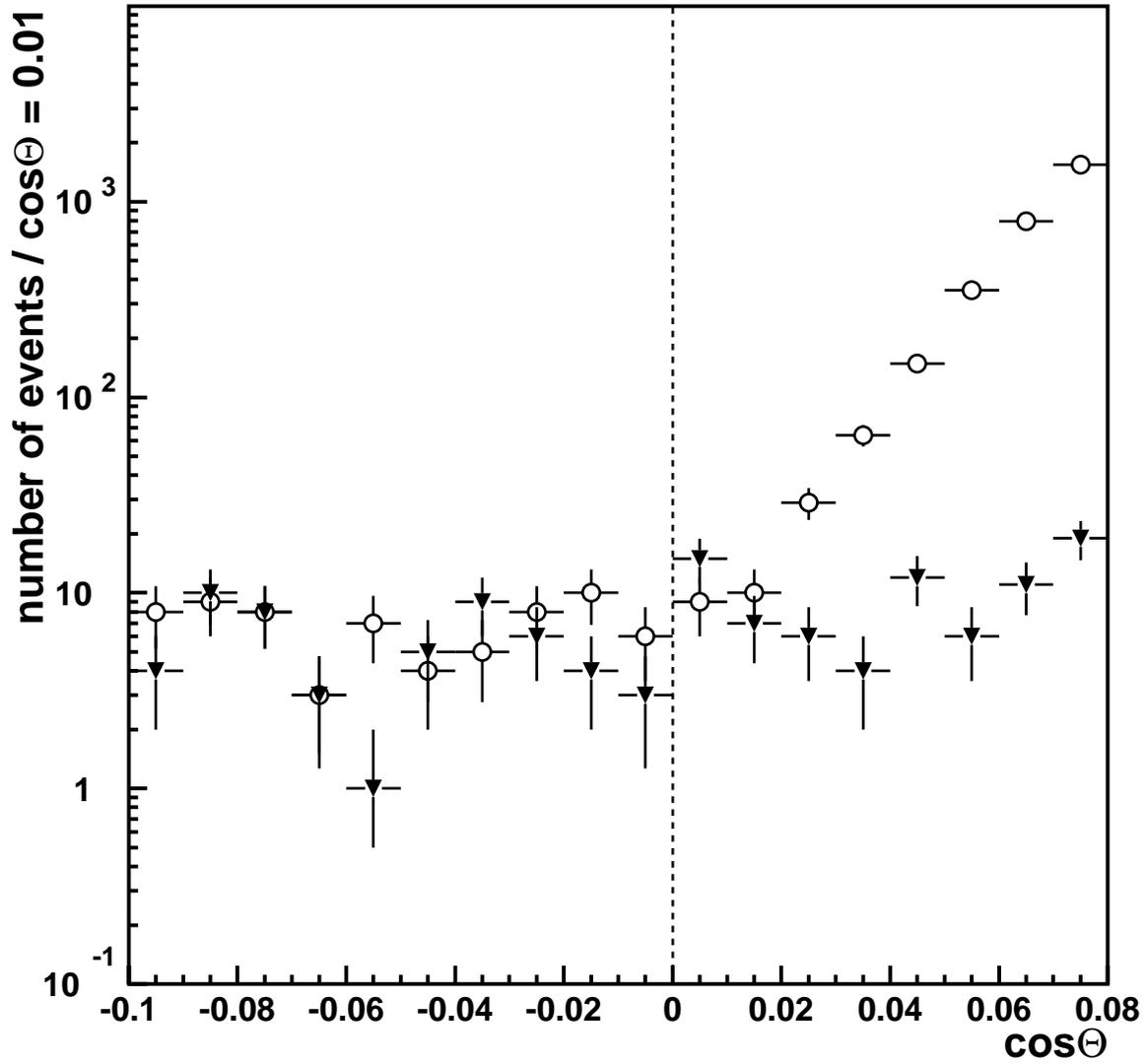}
  \caption{Zenith angle distribution 
    of through-going muons near the horizon observed by Super-K.  Filled
    triangles (open circles) indicate events coming from the 180$^\circ$
    azimuthal region where the rock overburden is thick (shallow).
    Most of the downward-going ($\cos \Theta >0$) muons denoted by
    filled triangles are induced by atmospheric neutrinos.}
  \label{skbg}
\end{figure}

Due to the finite fitter resolution and multiple Coulomb scattering in
the nearby rock, some down-going cosmic ray muons may appear to have
$\cos\Theta<0$.  Figure~\ref{skbg} illustrates the estimation of this
contamination.  Assuming this background continues to fall exponentially
as $\cos\Theta$ decreases, the contribution to apparent upward-going
muons is estimated to be 4.3$\pm0.4$ events, all contained in the $-0.1
< \cos\Theta < 0$ zenith angle bin.  The contamination at the Kamioka
site due to cosmic ray photoproduced upward-going pions\cite{up-pions}
meeting the 7m track length requirement is estimated to be \(<\) 1\%.

The total detection efficiency of the complete data reduction process
for upward through-going muons is estimated by a Monte Carlo simulation
to be \(>\)99\% which is almost isotropic for $-1 < \cos\Theta < 0$.
Using the upward/downward symmetry of the detector configuration, the
validity of this Monte Carlo program has been checked by real cosmic ray
downward through-going muons.

This analysis used a model which is a combination of the Bartol
atmospheric neutrino flux model~\cite{bartol} and a neutrino interaction
model composed of quasi-elastic scattering~\cite{qe} + single-pion
production~\cite{sp} + deep inelastic scattering (DIS) multi-pion
production.  The DIS cross-section is based on the parton distribution
functions (PDF) of GRV94DIS~\cite{grv94} with the additional kinematic
constraint of $W>1.4$~GeV/c$^2$.  Lohmann's muon energy loss formula in
standard rock~\cite{lohmann} is then employed to analytically calculate
the expected muon flux at the detector.  This flux is compared to three
other analytic calculations to estimate the model-dependent
uncertainties of the expected muon flux.  The other flux calculations
use the various pairs of the Bartol flux, the GRV94DIS PDF, the
atmospheric neutrino flux model calculated by Honda {\it et
  al}~\cite{honda}, and the CTEQ3M~\cite{cteq}PDF.  These comparisons
yield $\pm$10\% of uncertainty for the absolute flux normalization and
$-$3.7\% to +1.6\% for the bin-by-bin shape difference in the
zenith-angle distribution.  The shape difference is due mostly to the
input flux models.

The Bartol+GRV94DIS calculation results in an expected muon flux
$\Phi_{theo}$ of $(1.97\pm0.44(\rm{theo.}))\times10^{-13}{\rm{cm^{-2}}}$
${\rm{s^{-1}sr^{-1}}}$ ($\cos\Theta < 0$), where the estimated
theoretical uncertainties are described in
Table~\ref{systematictable_th}.  The dominant error comes from the
absolute normalization uncertainty in the neutrino flux, which is
estimated to be approximately $\pm20$\%~\cite{bartol,honda,frati} for
neutrino energies above several GeV.

Given the detector live time $T$, the effective area for upward
through-going muons $S(\Theta)$, and the detection efficiency
$\varepsilon(\Theta)$, the upward through-going muon flux is calculated
by the formula:
\begin{displaymath}
  \Phi_{obs}=\sum^{N}_{j=1} \frac{1}{\varepsilon(\Theta_j)}
  \cdot\frac{1}{S(\Theta_j)\,2\pi}\cdot\frac{1}{T}
\end{displaymath}
\noindent
where the suffix $j$ represents each event number, $2\pi$ is the total
solid angle covered by the detector for upward through-going muons, $N$
corresponds to the total number of observed muon events (614).
Subsequently, we subtract the cosmic ray muon contamination (4.3 events)
from the most horizontal bin (-0.1\(<\)$\cos\Theta$\(<\)0).

Conceivable experimental systematic errors are summarized in
Table~\ref{systematictable_ex}.  Including these experimental systematic
errors,  the observed upward through-going muon flux is:
$\Phi_{obs}=(1.74\pm0.07 {\mbox{(stat.)}}\pm0.02{\mbox{(sys.)}})
  \times10^{-13}{\rm{cm^{-2}s^{-1}sr^{-1}}}$.

\begin{figure}[thbp]
  \includegraphics[width=6.5in]{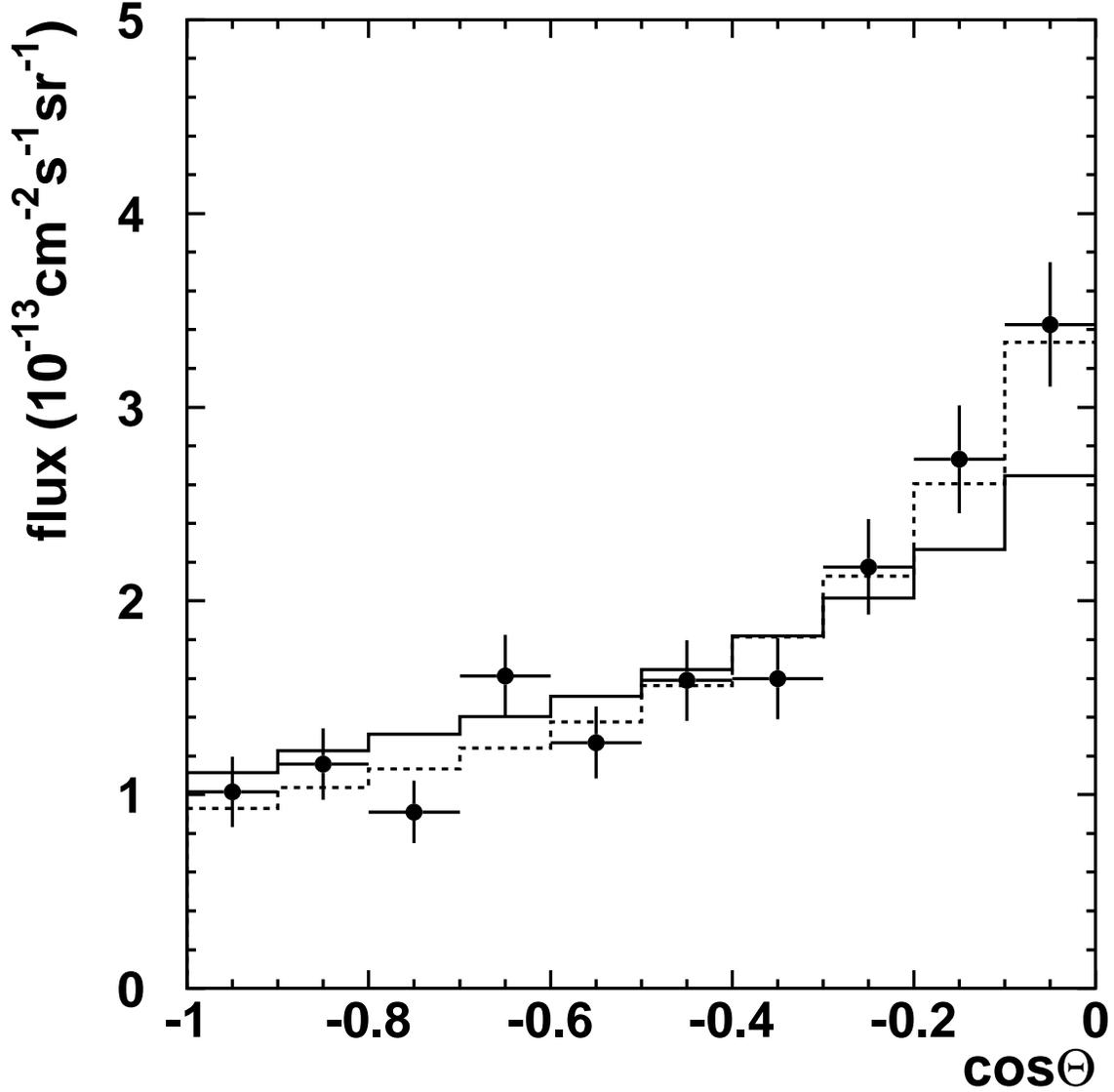}
  \caption{Upward through-going muon flux observed in Super-K
    as a function of zenith angle.  The error bars indicate uncorrelated
    experimental systematic plus statistical errors added in quadrature.
    The solid histogram shows the expected upward through-going muon
    flux with normalization ($\alpha_\mu=-14\%$) based on the Bartol
    neutrino flux for the null neutrino oscillation case.  Also shown as
    a dotted line is the expected flux assuming the best fit parameters
    at $(\sin^2 2\theta, \Delta m^2)=(0.95,
    5.9\times10^{-3}{\rm{eV}}^{2})$, $\alpha_\mu=+12\%$ for the
    $\nu_{\mu}\leftrightarrow\nu_{\tau}$ oscillation case.}
  \label{skangdist}
\end{figure}

Fig.~\ref{skangdist} shows the flux as a function of zenith angle.  The
shape of the distribution is not well represented by the theoretical
prediction without neutrino oscillation, having a $\chi^{2}$/degrees of
freedom (dof) = 18.7/9 corresponding to 2.8\% probability.  This shape
comparison is done after multiplying the expected flux by a free-running
normalization factor (1+$\alpha_\mu$), whose best fit value is
$\alpha_\mu=-14\%$.

A set of neutrino oscillation hypotheses are then tested using the
zenith angle distribution.  The expected flux ($(d\Phi/d\Omega)_{osc}$)
for a given set of $\Delta m^{2}$ and $\sin^{2}2\theta$ is calculated
and the same binning (dcos$\Theta$=0.1) is applied to this flux as to
the data.  To test the validity of a given oscillation hypothesis, we
minimize a $\chi^{2}$ which is defined as:
\begin{displaymath}
  \sum_{i=1}^{10}\left(\frac
  {\left(\frac{d\Phi}{d\Omega}\right)_{obs}^{i}-
    (1+\alpha_\mu)\left(\frac{d\Phi}{d\Omega}\right)_{osc}^{i}}
  {\sqrt{\sigma_{stat,i}^{2}+\sigma_{sys,i}^{2}}}\right)^{2}
   + \left(\frac{\alpha_\mu}{\sigma_{\alpha_\mu}}\right)^{2},
\end{displaymath}
\noindent
where $\sigma_{stat,i}$ ($\sigma_{sys,i}$) is the statistical
(experimental systematic) error in the observed flux
$(d\Phi/d\Omega)_{obs}^{i}$ for the $i$th bin, and $(1+\alpha_\mu)$ is
an absolute normalization factor of the expected flux.  The absolute
flux normalization error $\sigma_{\alpha_\mu}$ is estimated to be
$\pm$22~\% by adding in quadrature the bin-to-bin correlated
experimental errors and theoretical uncertainties in
Table~\ref{systematictable_th}.  Based on the bin-by-bin correlated
systematic errors in Table~\ref{systematictable_ex} added in quadrature,
we estimate $\sigma_{sys,i}$ to range from $\pm(0.3-3.8)$\%.  Then, the
minimum $\chi^{2} (\chi^{2}_{min})$ is searched for on the $\Delta
m^{2}-\sin^{2}2\theta$ plane.

\begin{figure}[thbp]
  \includegraphics[width=5.5in]{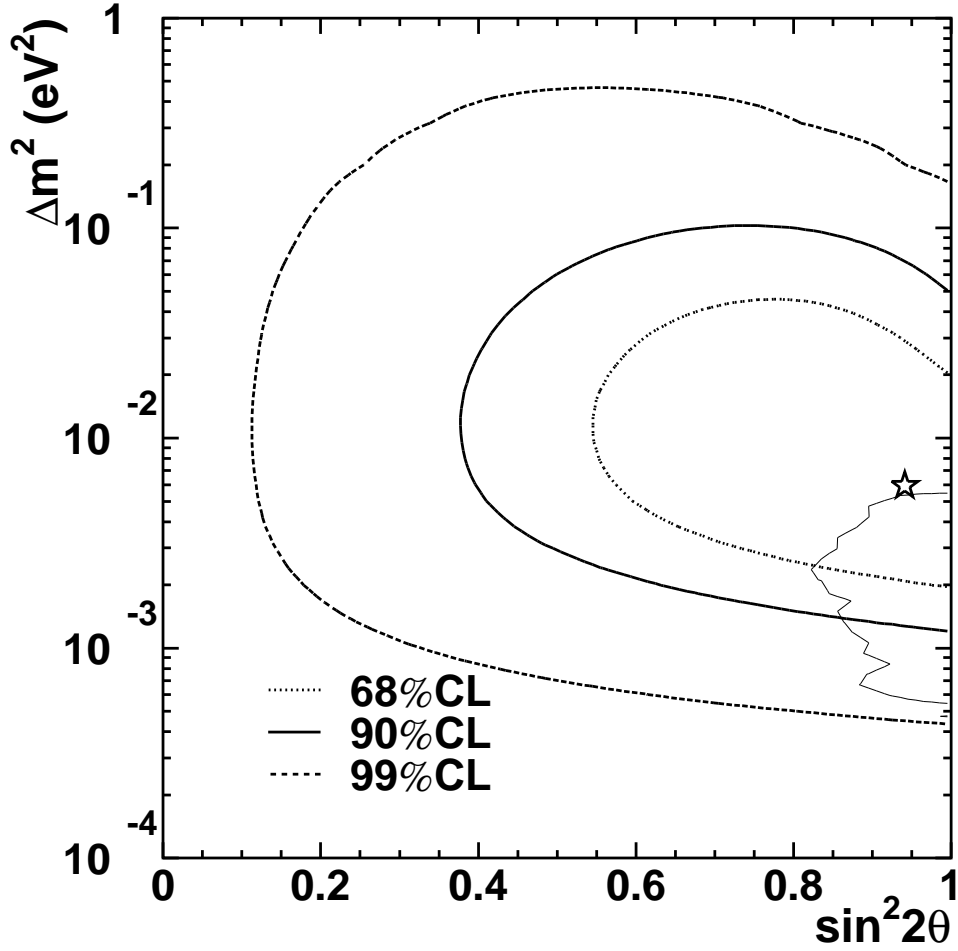}
  \caption{Allowed region contours at 68\% (dotted contour), 
    90\% (thick solid), and 99\% (dashed) C.L. obtained by the Super-K
    upward through-going muon analysis on the
    ($\sin^{2}2\theta$,$\Delta{m}^{2}$) plane for the
    $\nu_{\mu}\leftrightarrow\nu_{\tau}$ oscillation hypothesis.  The
    star indicates the best fit point at $(\sin^2 2\theta, \Delta
    m^2)=(0.95, 5.9\times 10^{-3}{\rm{eV}^2})$.  Also shown is the
    allowed region contour (thin solid) at 90\% C.L. by the Super-K
    contained event analysis.  The allowed regions are to the right of
    the contours.}
  \label{sknmntcontour}
\end{figure}

Assuming $\nu_{\mu}\leftrightarrow\nu_{\tau}$ oscillations,
$\chi^{2}_{min}(=7.5/8$ dof) occurs at $(\sin^2 2\theta, \Delta m^2)=
(0.95, 5.9\times 10^{-3}{\rm{eV}^2})$ and $\alpha_\mu=+12\%$, in good
agreement with the overall normalization found in the contained event
analysis~\cite{skosc}, although the $\alpha_\mu$ of this analysis refers
to the flux normalization of neutrino energies predominantly around
100~GeV.  For the null oscillation case ($\sin^2 2\theta$=0), we obtain
$\chi^{2}$ of 19.2 at a best-fit $\alpha_\mu=-14\%$ using the same
$\chi^{2}$ definition.  The zenith angle distribution of
$(1+\alpha_\mu)(d\Phi/d\Omega)^{i}_{osc}$ for the best fit parameters is
shown in Fig.~\ref{skangdist} together with the data.
Figure~\ref{sknmntcontour} shows the confidence intevals on the $(\sin^2
2\theta, \Delta m^2)$ plane for $\nu_{\mu}\leftrightarrow\nu_{\tau}$
oscillations.  The 90\%~C.L. contour marks the line of
$\chi^2_{min}+4.6$.  If we replace the Bartol neutrino
flux~\cite{bartol} by Honda's~\cite{honda} and/or the GRV94DIS parton
distribution functions~\cite{grv94} by CTEQ3M~\cite{cteq}, the allowed
region contours are similar to those presented in
Fig.~\ref{sknmntcontour}.  Consequently, we find that the zenith angle
dependence is in favor of the $\nu_{\mu}\leftrightarrow\nu_{\tau}$
oscillation hypothesis and supports the Super-K contained event
analysis~\cite{sksubgev,skmultigev,skosc}.  It is also consistent with
the data presented in the Kamiokande~\cite{kamupthrumu} and
MACRO~\cite{macro} upward-going muon analyses.  Interactions of
$\nu_\tau$ in the rock below is estimated at less than a few percent and
neglected in this analysis.  Oscillation of $\nu_\mu$ to $\nu_e$ in this
range of parameter space has been ruled out by the CHOOZ
experiment\cite{chooz}.

In conclusion, based on 614 upward through-going muon events during 537
detector live days, the flux of the upward through-going muons
(\(>\)1.6~GeV) is measured with the Super-K detector:
$\Phi_{obs}=(1.74\pm0.07 {\mbox{(stat.)}}\pm0.02
{\mbox{(sys.)}})\times10^{-13} {\rm{cm}^{-2}{s}^{-1}{sr}^{-1}}$.  This
is compared with the expected flux of
$\Phi_{theo}=(1.97\pm0.44(\rm{theo}.))\times10^{-13}
{\rm{cm}^{-2}{s}^{-1}{sr}^{-1}}$.  The absolute observed upward
through-going muon flux is in agreement with the expected flux within
the relatively large uncertainties in the theoretical calculations.  We
find that the zenith angle dependence does not agree with the
theoretical expectation without neutrino oscillations at the 97\% C.L.
However, the $\nu_{\mu}\leftrightarrow\nu_{\tau}$ oscillation hypothesis
with $\sin^22\theta > 0.4$ and $1\times 10^{-3} < \Delta m^2 < 1\times
10^{-1}$ eV$^{2}$ is consistent with the observed zenith angle shape at
90\%~C.L.  This result supports the evidence for neutrino oscillations
given by the analysis of the contained atmospheric neutrino events by
Super-K.

We gratefully acknowledge the cooperation of the Kamioka Mining and
Smelting Company.  The Super-Kamiokande experiment has been built and
operated from funding by the Japanese Ministry of Education, Science,
Sports and Culture, and the United States Department of Energy.

\begin{table}[tbhp]
  \caption{List of theoretical uncertainties in the flux calculation.}
  \begin{tabular}{cl|c}
    \multicolumn{2}{c|}{Error source} & Error (\%) \\
    \hline
    \multicolumn{2}{l|}{Chemical composition of the rock} 
                       & $\ll$1\tablenotemark[1]\\ 
    \multicolumn{2}{l|}{$\nu$ flux normalization}
                       & $\pm$20\tablenotemark[1] \\ \hline
    \multicolumn{2}{l|}{Theoretical model dependence} & \\
                       & absolute flux & $\pm$10\tablenotemark[1] \\
                       & bin by bin & -3.7 to +1.6\tablenotemark[2]\\
                       & spectral index & $\pm$1.4\tablenotemark[1]\\
  \end{tabular}
  \tablenotetext[1]{Theoretical bin-by-bin uncertainty}
  \tablenotetext[2]{Theoretical uncorrelated correlated uncertainty}
  \label{systematictable_th}
\end{table}

\begin{table}[thbp]
  \caption{List of experimental systematic errors in the flux
    measurement.}
  \begin{tabular}{l|l}
    Error source & Error (\%) \\
    \hline
    Uncertainty in $\Delta\theta_{rec}$ & $<\pm$1\tablenotemark[1]\\
    Detection efficiency                & $< \pm 1.2$\tablenotemark[2]\\
    7m track length cut                 & $\pm$0.5\tablenotemark[3]\\
    Live time                           & $\pm$0.1\tablenotemark[3]\\
    Effective area                      & $\pm$0.3\tablenotemark[3]\\
    PMT gain                            & $\ll$1\tablenotemark[3]\\
    Water transparency                  & $\ll$1\tablenotemark[3]\\
  \end{tabular}
  \tablenotetext[1]{Experimental uncorrelated systematic error
    specific in the most horizontal bin $-$0.1\(<\)cos$\Theta$\(<\)0}
  \tablenotetext[2]{Experimental uncorrelated systematic error}
  \tablenotetext[3]{Bin-by-bin correlated experimental systematic errors}
  \label{systematictable_ex}
\end{table}

\end{document}